# Engineering Methods for Differentially Private Histograms: Efficiency Beyond Utility

Georgios Kellaris
Harvard University & BU
gkellaris@seas.harvard.edu

Stavros Papadopoulos
Intel Labs & MIT
stavrosp@csail.mit.edu

Dimitris Papadias
HKUST
dimitris@cs.ust.hk

## ABSTRACT

Publishing histograms with $\epsilon$-*differential privacy* has been studied extensively in the literature. Existing schemes aim at maximizing the *utility* of the published data, while previous experimental evaluations analyze the privacy/utility trade-off. In this paper we provide the first experimental evaluation of differentially private methods that goes beyond utility, emphasizing also on another important aspect, namely *efficiency*. Towards this end, we first observe that all existing schemes are comprised of a small set of common blocks. We then optimize and choose the best implementation for each block, determine the combinations of blocks that capture the entire literature, and propose novel block combinations. We qualitatively assess the quality of the schemes based on the skyline of efficiency and utility, i.e., based on whether a method is dominated on both aspects or not. Using exhaustive experiments on four real datasets with different characteristics, we conclude that there are always trade-offs in terms of utility and efficiency. We demonstrate that the schemes derived from our novel block combinations provide the best trade-offs for time critical applications. Our work can serve as a guide to help practitioners *engineer* a differentially private histogram scheme depending on their application requirements.

## 1. INTRODUCTION

The histogram is a basic statistical tool for describing a dataset distribution. Publishing histograms raises privacy concerns, which motivated numerous works [1, 24, 26, 17, 5, 14, 21, 23] to target at preserving privacy through the concept of $\epsilon$-*differential privacy* [9]. The idea is to perturb the histogram bins prior to their publication in a way that protects each individual record in the data. The $\epsilon$ parameter controls the privacy level. Lower $\epsilon$ values offer better privacy, but increase the perturbation. The higher the perturbation, the higher the error and, thus, the lower the *utility* of the published histogram.

Existing work on differentially private histograms focuses solely on maximizing data utility, for a given $\epsilon$ and for various query types. In addition, previous experimental evaluations on the topic [21, 13] assess the quality of a method based solely on the trade-off between privacy and utility for various datasets. We argue that *efficiency* is an important aspect of differentially private histogram methods that has not been explored.

Efficiency refers to the time complexity of a method, which depends on the histogram size. In many practical applications, the published histogram is desirable to have fine granularity and, hence, large size; examples include time-series (e.g., where the bins are time intervals in traffic reporting), demographics (e.g., where the bins represent annual income), or medical data (e.g., where the bins represent blood sugar levels). The current state-of-the-art differentially private histogram scheme, DAWA [17], runs in time $O(N^3 \log N)$ for histogram size $N$ when considering arbitrary histogram queries. For histogram sizes in the orders of tens of thousands (such as in the examples above), DAWA would take months to complete. Although one may argue that this is an offline task, it would still consume significant resources. Moreover, in a streaming setting, histograms need to be published periodically and, thus, efficiency is crucial for the application.

In this paper we provide the first experimental evaluation of differentially private histogram methods that goes beyond utility to also emphasize efficiency. Toward this end, we follow a novel approach as compared to [21, 13]. These works consider each existing method as a black box and carry out a comparison only in terms of utility. In contrast, we first observe that there is a small set of common components across all methods, and that each method essentially proposes a different instantiation of each component. Then we identify the best algorithm for each component, and propose variants that outperform the existing ones in at least one aspect (efficiency or utility). We show that the entire bibliography can be captured with combinations of the outlined components. Finally, we develop novel combinations that do not appear in the literature.

We provide extensive experiments over four real datasets with different characteristics. The main takeaway is that there is no best solution, but rather there are trade-offs on utility and efficiency. In order to qualitatively assess the methods, we calculate the *skyline* of efficiency versus utility; a method on a skyline is not worse than (or dominated by) another in terms of *both* the two examined aspects, which means that it is interesting in at least one of the aspects. We further demonstrate that the performance of each method greatly depends on the setting, i.e., query type, data characteristics, and available privacy budget. More importantly, we show that for time critical applications, our novel combinations dominate all other techniques. Effectively, our work shows how a practitioner can *engineer* a differentially private histogram method by combining different building blocks and assessing the quality based on the efficiency/utility skyline to suit her needs.

The remainder of the paper is organized as follows. Section 2 provides the necessary preliminaries. Section 3 surveys the related work. Section 4 introduces the basic components in differentially private histogram methods. Section 5 describes our proposed component optimizations. Section 6 introduces mechanisms based on novel component combinations. Section 7 includes our exhaustive experimental evaluation. Section 8 concludes our work.

## 2. PRELIMINARIES

Let $\mathcal{D}$ be a collection of datasets. We define a family of functions $\mathcal{F} = \{F_j : \mathcal{D} \rightarrow \mathcal{H}\}$, such that for all $j$ and all $D \in \mathcal{D}$, $F_j(D) = \mathbf{h} \in \mathcal{H}$ is an (ordered) vector called *histogram*. An element of $\mathbf{h}$ is termed *bin* and consists of a value and a label, where $\mathbf{h}[i]$ represents the $i^{\text{th}}$ bin value of $\mathbf{h}$. All histograms have the property that any record in $D$ increments *at most a single* $\mathbf{h}[i]$ by 1. Finally, we call $F_j$ a *histogram algorithm*. For instance, let $D \in \mathcal{D}$ be a dataset of medical records. Then, $F_1 \in \mathcal{F}$ may produce histogram $\mathbf{h}_1$ such that $\mathbf{h}_1[i]$ is the number of patients in $D$ having age $i$, and $F_2 \in \mathcal{F}$ may produce histogram $\mathbf{h}_2$ such that $\mathbf{h}_2[i]$ is the number of patients in hospital with id $i$. Observe that the presence of a patient in $D$ increments at most one bin by 1 in both histograms.

Our goal is to publish a $N$-element histogram $\mathbf{h}$ produced by some *fixed* algorithm $F$ on a $D \in \mathcal{D}$, while satisfying $\epsilon$-*differential privacy* and allowing *arbitrary point and interval queries* on its bins with high utility. Specifically, we define an interval query as a range of bins $[i_l, i_u]$, $1 \leq i_l \leq i_u \leq N$, which returns the sum $\sum_{i=i_l}^{i_u} \mathbf{h}[i]$. In our example above, an interval query on $\mathbf{h}_1$ could be $[10, 20]$, asking for the number of patients between 10 and 20 years old. Point queries degenerate from the interval queries when $i_l = i_u$. We assume that the queries are *not known* prior to the publication of the histogram.

To achieve $\epsilon$-differential privacy, we apply a mechanism $M$ on the histogram, which perturbs it in a way that satisfies the following definition, adapted from [6].

Definition 1. *A mechanism* $M : \mathcal{H} \rightarrow \hat{\mathcal{H}}$ *satisfies* **$\epsilon$-differential privacy** *for a histogram algorithm* $F \in \mathcal{F}$, *if for all sets* $\hat{H} \subseteq \hat{\mathcal{H}}$, *and every pair* $D, D' \in \mathcal{D}$ *where* $D'$ *is obtained from* $D$ *by removing a record* ($D, D'$ *are called neighboring), it holds that*

$$\Pr[M(F(D)) \in \hat{H}] \leq e^{\epsilon} \cdot \Pr[M(F(D')) \in \hat{H}]$$

Intuitively, $\epsilon$-differential privacy guarantees that the perturbed histogram $\hat{H}$ will be the same with high probability (tunable by $\epsilon$), regardless of whether a patient agrees to participate in the publication or not. Equivalently, the sensitive information of any patient cannot be inferred from the published data.

Definition 2. *The* **sensitivity** *of any histogram algorithm* $F \in \mathcal{F}$ *is* $\Delta(F) = \max_{D,D' \in \mathcal{D}} \|F(D) - F(D')\| = 1$ *for all neighboring* $D, D' \in \mathcal{D}$.

In other words, the sensitivity of $F$ represents how much the histogram $F(D)$ changes when a record is deleted from $D$. Since any record contributes 1 to at most a single bin, the sensitivity is 1 for any histogram algorithm $F \in \mathcal{F}$.

The most basic technique for achieving $\epsilon$-differential privacy adds Laplace noise to the histogram bins using the Laplace Perturbation Algorithm (LPA [9, 7]). Let $Lap(\lambda)$ be a random variable drawn from a Laplace distribution with mean zero and scale parameter $\lambda$. LPA achieves $\epsilon$-differential privacy through the mechanism outlined in the following theorem, adapted from [9].

Theorem 1. *Let* $F \in \mathcal{F}$ *and define* $\mathbf{h} \stackrel{\text{def}}{=} F(D)$. *A mechanism* $\mathcal{M}$ *that adds independently generated noise from a zero-mean Laplace distribution with scale parameter* $\lambda = \Delta(F)/\epsilon = 1/\epsilon$ *to each of the values of* $\mathbf{h}$, *i.e., which produces transcript* $\hat{\mathbf{h}} = \mathbf{h} + \langle Lap(1/\epsilon)\rangle^N$, *enjoys* $\epsilon$-*differential privacy.*

With LPA, an interval query $[i_l, i_u]$ is processed on the noisy $\hat{\mathbf{h}}$ and returns $\sum_{i=i_l}^{i_u} \hat{\mathbf{h}}[i]$. The Laplace noise injected in each bin introduces error, which is aggregated when the noisy bin values are added. For large ranges, this error may completely destroy the utility of the answer. Numerous works (overviewed in Section 3) introduce alternative mechanisms for improving the utility of the output histograms in the case of interval queries.

Next, we include a useful *composition* theorem (adapted from [19]) based on [9, 8]. It concerns executions of multiple differentially private mechanisms on non-disjoint and disjoint inputs.

Theorem 2. *Let* $M_1, \ldots, M_r$ *be mechanisms, such that each* $M_i$ *provides* $\epsilon_i$-*differential privacy. Let* $\mathbf{h}_1, \ldots, \mathbf{h_r} \in \mathcal{H}$ *be histograms created on pairwise non-disjoint (resp. disjoint) datasets* $D_1, \ldots, D_r$, *respectively. Let* $M$ *be another mechanism that executes* $M_1(\mathbf{h}_1), \ldots, M_r(\mathbf{h}_r)$ *using independent randomness for each* $M_i$, *and returns their outputs. Then,* $M$ *satisfies* $\left(\sum_{i=1}^{r} \epsilon_i\right)$-*differential privacy (resp.* $(\max_{i=1}^{r} \epsilon_i)$-*differential privacy).*

The above theorem allows us to view $\epsilon$ as a *privacy budget* that is distributed among the $r$ mechanisms. Moreover, note that the theorem holds even when $M_i$ receives as input the private outputs of $M_1, \ldots, M_{i-1}$ [19].

## 3. RELATED WORK

Existing literature on differentially private histograms aims at improving upon LPA in terms of utility. While there are theoretical lower bounds [3, 10, 2, 4] on the utility of differentially private mechanisms for point and interval queries, different methods offer different utility in practice. The methods can be divided into two categories; *data-aware* (Section 3.1) that utilize *smoothing*, and *data-oblivious* (Section 3.2) that rely on *hierarchical* tree structures. In Section 3.3 we discuss the performance of methods in each category and previous experimental evaluations on the topic.

## 3.1 Data-aware Methods

These approaches first smooth the histogram, typically either by grouping similar bin values and substituting them with their average, or by performing a smoothing filter such as the Discrete Fourier Transform (DFT). Then, they apply Laplace noise similar to LPA to the averages or the DFT coefficients. Interval queries are processed by summing the histogram bin values in the range. As such, the output error increases linearly with the interval length. Smoothing reduces the sensitivity and, hence, the injected Laplace noise, but adds approximation error. As a result, smoothing methods are effective if the Laplace noise error reduction exceeds the smoothing approximation error. The bin grouping algorithm assigns scores to a set of potential grouping strategies, and selects the one with the minimum score, in a manner that does not compromise differential privacy. Existing approaches differ in the set of examined strategies, the scoring function, and the selection process.

The SF algorithm [24] follows the grouping and averaging paradigm. Specifically, given as input a *fixed* parameter $k$ and privacy budgets $\epsilon$, $\epsilon'$, SF initially finds a set of $k$ groups of *contiguous* bins through an $\epsilon'$-differentially private process. Subsequently, it smooths the bin values based on the grouping, and adds Laplace noise generating $(\epsilon - \epsilon')$-differentially private histogram. Due to linear composition (Theorem 2), the SF mechanism achieves $\epsilon$-differential privacy. The grouping sub mechanism of SF operates on the original histogram and determines the $k$ groups such that the estimated *squared* error is minimized. This error is expressed as the sum of (i) the squared approximation error due to smoothing, and (ii) the squared error from injecting Laplace noise with scale $1/(\epsilon - \epsilon')$ prior to publication. It then applies the exponential mechanism [20] in order to alter the group borders and achieve $\epsilon'$-differential privacy. Note that, due to this step, the total error of SF deviates from the actual minimum. The grouping component of SF runs in $O(N^2)$, where $N$ is the number of histogram bins.

Acs et al. [1] present two mechanisms, EFPA and P-HP. EFPA is an improvement of [22], which smooths the histogram using a subset of its DFT coefficients perturbed with Laplace noise, while guaranteeing that the output histogram satisfies $\epsilon$-differential privacy. P-HP is a grouping and averaging method that improves SF [24]. In particular, instead of receiving the number of groups $k$ as input, it discovers the optimal value of $k$ on-the-fly. Contrary to SF, it utilizes an *absolute* error metric. The grouping algorithm of P-HP runs also in $O(N^2)$, but similarly to SF does not examine all possible groups. P-HP is shown to outperform both EFPA and SF in terms of utility [1].

AHP [26] first applies LPA to the histogram with noise scale $1/\epsilon'$, and *sorts* the resulting bins in descending order. Subsequently, it executes a grouping and averaging technique that is different from SF and P-HP. Specifically, it operates on already $\epsilon'$-differentially private data and, hence, does not need to apply the exponential mechanism. Moreover, it finds the grouping that minimizes the *squared* error metric expressed as a function of the noisy data, rather than the original histogram (and, thus, similar to [24, 1], it does not guarantee the actual minimum error). Note that the ordering attempts to minimize the approximation error, since it results in groups with more uniform bin values. The authors present two algorithms; one that evaluates all possible groups and runs in $O(N^3)$ time, and a greedy one that considers only a subset of the possible options and runs in $O(N^2)$. They conducted experiments using the latter, and demonstrated that AHP offers better utility than P-HP.

DAWA [17] comprises of two stages. The first stage executes a smoothing technique, while the second an optimized version of the matrix mechanism [18]. Its grouping and averaging component invests $\epsilon'$ budget to reduce the *absolute* error metric similar to [1]. However, instead of executing the exponential mechanism, it adds noise to the costs of the groups used in the selection process on-the-fly. The authors present two instantiations; the first evaluates all possible groupings and runs in $O(N^2 \log N)$ time, whereas the second considers only a subset and runs in $O(N \log^2 N)$. The output of the smoothing procedure is fed to the matrix mechanism. The latter belongs to a category of schemes [18, 25, 12] that take as input a set $\mathbf{W}$ of *pre-defined* interval queries, and assign more privacy budget to the bins affecting numerous queries. The total time complexity of DAWA is $O(|\mathbf{W}|N \log N)$. Finally, DAWA can be adapted to our setting of *arbitrary* queries in two ways; either by completely ignoring the second stage, resulting in time complexity $O(N^2 \log N)$ (or $O(N \log^2 N)$ in the approximate version), or by feeding all the possible queries to the input of the matrix mechanism, yielding time complexity $O(N^3 \log N)$.

## 3.2 Data-oblivious Methods

These are hierarchical schemes that build an aggregate tree on the original histogram; each bin value is a leaf, and each internal node represents the sum of the leaves in its subtree. In order to achieve $\epsilon$-differential privacy, they add Laplace noise to each node, which is proportional to the tree level (since each bin value is incorporated in all the sums along its path to the root). An interval query is processed by identifying the maximal subtrees that exactly cover the interval, and summing the values stored in their roots. Compared to LPA, these methods essentially increase the sensitivity from 1 to $\log N$, but sum fewer noisy values when processing the interval query, reducing the aggregate error. Essentially, they render the output error polylogarithmic to the number of bins $N$ and independent of the range query length. Moreover, their time complexity is $O(N)$.

Hay et al. [14] build a binary aggregate tree and inject Laplace noise uniformly across all nodes. In addition to constructing the final interval from the roots of the maximal subtrees that cover the interval, they also explore other node combinations. Independently from [14], Privelet [23] builds a Haar wavelet tree and adds Laplace noise, achieving practically the same effect as [14]. Based on the observation that the privacy budget should not be divided equally among all levels, Cormode et al. [5] enhance [14] with a geometric budget allocation technique. Qardaji et al. [21] survey the above approaches, concluding that the theoretical optimal fan-out of the tree is 16. They experimentally show that [14], when combined with the budget allocation of [5] and their optimal fan-out, outperforms Privelet and SF.

## 3.3 Discussion

Smoothing techniques (with or without ordering) aim at point and short interval queries because their error increases linearly with the number of histogram bins within the range query. Moreover, they can adapt to limited privacy budgets by generating a small number of large groups. The additional matrix mechanism in DAWA improves the utility for pre-defined ranges of any length. Data oblivious (i.e., hierarchical) methods are designed mostly for long interval queries, and their utility is independent of the underlying data distribution. However, they split the privacy budget among the tree levels or input queries, and as such, they perform better with higher $\epsilon$ values.

There are two experimental evaluations on differentially private histograms. Qardaji et al. [21] compare all data-oblivious methods, and determine which optimizations (e.g., splitting the budget unevenly among the tree levels or using multiple tree node combinations to compute an interval query) offer the best experimental results in terms of utility. Hay et al. [13] assess existing methods in terms of output error, for different input data sizes, distributions, and query workloads. They mainly focus on how well the data-aware methods adapt to $\epsilon$ and the number of input records $n$.

In our work, we go *beyond* utility and also include computational *efficiency*. More importantly, we do not use the existing methods as black boxes. Instead, we decompose each method into algorithmic components. For example, SF [24] determines the consecutive histogram bins that minimize the *average squared error* when they are smoothed. P-HP [1] implements the same component, but focuses on the *average absolute error*. Similarly, both [14] and [5] build an aggregate tree over the histogram bins, but they allocate the privacy budget differently among the tree levels. Furthermore, we observe that each component can be instantiated differently. For instance, smoothing using the average squared error can have two different implementations [24, 26]. In our evaluation, we determine the best instantiation for each component (i.e., the one that is superior in terms of both utility and time complexity). In addition, we propose new component implementations and devise novel component combinations.

## 4. BASIC COMPONENTS

In this section, we explain how all the existing techniques can be expressed as different combinations of four basic components: *Smoothing*, *Ordering*, *Fixed Queries*, and *Hierarchical*. For each component We identify the best algorithm in terms of utility and time complexity.

*Smoothing.* All data-aware mechanisms [24, 1, 26, 17] of Section 3 utilize smoothing. Initially, smoothing spends budget $\epsilon_1$ to discover the groups (denoted as $g_i$) for the input histogram $\mathbf{h}$ that minimize the output error. The tasks performed in order to effectively determine the groups are elaborated further below. Then, it groups the bins of $\mathbf{h}$ according to the computed groups, and averages their values. Next, it adds noise to the respective average with scale $1/(\epsilon_2 \cdot |g_i|)$. Finally, it sets the noisy average of every group $g_i$ as the value of the bins in $g_i$, and outputs the noisy smoothed histogram $\hat{\mathbf{h}}$.

The grouping procedure of smoothing determines the way the bins are privately grouped. In all methods, this is modeled as an optimization problem where the resulting grouping must minimize a certain error metric. Specifically, grouping takes as input a histogram $\mathbf{h}$, privacy budgets $\epsilon_1$ and $\epsilon_2$, and an error metric $\mu$. Its goal is to find the groups that minimize $\mu$, while satisfying $\epsilon_1$-differential privacy. Let $G$ be a *grouping strategy*, i.e., a set of $|G|$ groups of *contiguous* bins that cover all histogram bins and are mutually disjoint. Let $b_j$ be a bin value, and $\bar{g}_i$ the average of the bins in group $g_i \in G$, i.e., $\bar{g}_i = \sum_{b_j \in g_i} b_j / |g_i|$.

The total error has two components. The first is due to the smoothing process and depends on the difference between the value $b_j$ of a bin and the average $\bar{g}_i$ of the group in which it belongs. The second component is due to the noise injected after the grouping and averaging by utilizing budget $\epsilon_2$. The *absolute* and *squared* error metrics combine the two components in different ways. Both metrics represent the collective error per bin, rather than the final error in an interval query. Thus, smoothing mainly aims at maximizing the accuracy of point queries.

**Absolute error.** This metric is defined in [1, 17] as:

$$err_1 = \sum_{i=1}^{|G|} \left( \sum_{b_j \in g_i} |b_j - \bar{g}_i| + \frac{1}{\epsilon_2} \right) \tag{1}$$

The state-of-the-art algorithm that uses the absolute error is the smoothing component of DAWA [17], which works as follows. Initially, it calculates the cost $c_i = \sum_{b_j \in g_i} |b_j - \bar{g}_i| + \frac{1}{\epsilon_2}$ of each group $g_i$ in Equation 1 by utilizing a binary search tree in $O(\log N)$ time. Then, it adds noise with scale $1/(\epsilon_1 |g_i|)$ to $c_i$ producing $\hat{c}_i = \sum_{b_j \in g_i} |b_j - \bar{g}_i| + \frac{1}{\epsilon_2} + Lap(1/(\epsilon_1 |g_i|))$. Finally, it finds the groups that minimize $\hat{err}_1 = \sum_{i=1}^{|G|} \hat{c}_i$ using dynamic programming in $O(N^2)$ time. The total time of grouping is dominated by that of computing the costs of all the $O(N^2)$ groups, which is $O(N^2 \log N)$.

**Squared error.** This metric is defined in [24, 26] as:

$$err_2 = \sum_{i=1}^{|G|} \left( \sum_{b_j \in g_i} (b_j - \bar{g}_i)^2 + \frac{1}{\epsilon_2^2} \right) \tag{2}$$

The state-of-the-art grouping algorithm that utilizes the squared error is AHP [26], which works as follows. It adds noise with scale $1/\epsilon_1$ to each bin of the initial histogram, and computes cost $\hat{c}_i = \sum_{\hat{b}_j \in g_i} (\hat{b}_j - \bar{g}_i)^2 + \frac{1}{\epsilon_2^2}$, where $\hat{b}_j$ is a noisy bin value, and $\bar{g}_i$ the average of a group of noisy bins. Finally, it finds the groups that minimize $\hat{err}_2 = \sum_{i=1}^{|G|} \hat{c}_i$. Its time complexity is $O(N^3)$.

The smoothing component returns a noisy histogram, and the queries are answered by summing noisy bin values. As such, its error depends on the interval length linearly. The running time of the component is $N^2 \log N$ using the algorithm of [17] for the absolute error metric, and $N^3$ using the algorithm of [26] for the squared error metric. We reduce the running time to the optimal value for the squared error metric in Section 5.1. Moreover, [17] has presented an approximate version of smoothing that only considers groups whose size is a power of 2. The approximation has running time of $O(N \log N)$. Depending on whether we use this approximation or not, we can have three variations of smoothing: approximate smoothing, smoothing with absolute error metric, and smoothing with squared error metric.

*Ordering.* Ordering [16] can be applied before smoothing. It receives the histogram $\mathbf{h}$ and budget $\epsilon$, it adds (Laplace) noise with scale $\lambda = 1/\epsilon$ to each bin value, and sorts them in descending order. The running time of ordering is $O(N \log N)$, using any optimal sorting algorithm.

*Fixed Queries.* The Fixed Queries component is the building block of methods that target interval queries known a priori (DAWA [17]). It receives as input a histogram $\mathbf{h}$, a privacy budget $\epsilon$, and a query workload $\mathbf{W}$. It executes an off-the-shelf mechanism such as MWEM [12] or the matrix mechanism [18], and outputs the noisy histogram $\hat{\mathbf{h}}$. In this work, we use the optimized matrix mechanism algorithm of [17] for this component, which has the lowest time complexity (i.e., $O(|\mathbf{W}|N \log N)$) and comparable utility to the competitors. Note that in our settings, the queries are not known a priori. Thus, the fixed queries component works with workload all $O(N^2)$ possible queries. In this case, it degenerates to the aggregate tree method [17], and its running time becomes $O(N^3 \log N)$). In order to benefit from it, we should combine it with smoothing.

*Hierarchical.* The hierarchical component builds an aggregate tree and characterizes the data-oblivious methods. The best implementation is described in [21]. It receives as input a histogram $\mathbf{h}$, privacy budget $\epsilon$, and builds an aggregate tree of fan-out $f$[1] and height $t$. It splits the budget $\epsilon$ into $t$ budgets such that $\sum_{i=1}^{t} \epsilon_i = \epsilon$, and adds Laplace noise with scale $1/\epsilon_i$ to tree level $i$, for each $i$. In order to maximize utility, the hierarchical component answers queries by combining nodes from the noisy tree using the method of [14] (see Section 3). For an interval covering $m$ bins, this component induces $O(\log m \log N)$ error, as opposed to LPA that inflicts $O(m)$ error. Therefore, the hierarchical methods exhibit benefits for large intervals.

**"Decomposing" the Literature.** Some works build upon a single standalone component; [24, 1, 26] are essentially different instantiations of smoothing, whereas [14, 23, 5] implement the hierarchical component. On the other hand, others schemes constitute combinations of two components. Specifically, [17] initially applies smoothing with the absolute error metric or approximate smoothing, and then uses the fixed queries component. Finally, [26] utilizes the ordering component and then applies smoothing with the squared error metric. Identifying these basic components allows us to (i) carefully study and optimize them (Section 5), and (ii) construct new efficient and effective schemes via the seamless combination of these components (Section 6).

# 5. COMPONENT OPTIMIZATIONS

We optimize two basic components; (i) the smoothing with the squared error metric, resulting in better utility and running time of $O(N^2)$, which we also show that is optimal, (ii) the fixed queries by reducing its time complexity by a factor of $N$, while maintaining its utility.

## 5.1 Smoothing

We introduce an *optimal* way to compute the squared error, which (i) reduces the time complexity of the current best method by a factor of $N$, and (ii) improves the accuracy of smoothing.

The following theorem provides a lower bound on the time complexity of smoothing. The lower bound applies to *both* squared and absolute error metrics.

[1]We set $f = 16$ because it is optimal in terms of utility for interval queries [21].

THEOREM 3. *A grouping algorithm on a histogram with $N$ bins runs in $\Omega(N^2)$.*

PROOF. The number of all the possible groups is $\Theta(N^2)$. This is because we have $N$ groups of size 1, $N-1$ groups of size 2, and so on (recall that a permissible group can only consist of contiguous bins). Thus, the total number of groups is $N+(N-1)+(N-2)+\ldots+1 = \frac{N(N+1)}{2}$. It suffices to prove that there is an input for which any algorithm must check all the possible groups at least once.

We build a histogram such that every group $g_i$ contributes cost $\hat{c}_i = |g_i|$ (i.e., equal to its cardinality) to the error metric. In this scenario, *any* grouping strategy $G$ minimizes the error metric, since *every* $G$ leads to error $\sum_{g_i \in G} \hat{c}_i = N$. Now suppose that we reduce the cost of a random group $g_j$ to $|g_j| - \delta$ for some $\delta > 0$. Any grouping strategy that includes $g_j$ will result in error $N - \delta$, whereas any other will result in $N$. Therefore, the grouping strategy $G^*$ that minimizes the error metric *must* include $g_j$. Since $g_j$ is a random group, the algorithm that finds $G^*$ must check the $\hat{c}_i$ of *every* group $g_i$ in order to find $g_j$. □

We next present an algorithm that minimizes the squared error $\hat{err}_2$ in $O(N^2)$ time. Therefore, due to the lower bound in Theorem 3, our algorithm is *optimal*. Given that $\hat{g}_i = \frac{\sum_{\hat{b}_j \in g_i} \hat{b}_j}{|g_i|}$, we observe that the cost of each group can be rewritten as follows.

$$\hat{c}_i = \sum_{\hat{b}_j \in g_i} \left(\hat{b_j} - \bar{g}_i\right)^2 + \frac{1}{\epsilon_2^2} = \sum_{\hat{b}_j \in g_i} \hat{b_j}^2 - \frac{\left(\sum_{\hat{b}_j \in g_i} \hat{b}_j\right)^2}{|g_i|} + \frac{1}{\epsilon_2^2}$$

Based on the above equation, we can efficiently compute the cost of each group using the following procedure. Initially, we add noise with scale $1/\epsilon_1$ to every histogram bin. In a *pre-processing stage*, we build vector $\mathbf{v}_1$ that stores the noisy bin values $\hat{b}_j$, and vector $\mathbf{v}_2$ that stores their squares $\hat{b}_j^2$. Subsequently, we construct the *prefix sums* for each vector. Specifically, the prefix sums for $\mathbf{v}_1$ ($\mathbf{v}_2$) is a vector $\mathbf{v}'_1$ ($\mathbf{v}'_2$), such that $\mathbf{v}'_1[j] = \sum_{i=1}^{j} \mathbf{v}_1[i]$ ($\mathbf{v}'_2[j] = \sum_{i=1}^{j} \mathbf{v}_2[i]$). The pre-processing takes $O(N)$ time. For each group $g_i$ over contiguous bins $l, l+1, \ldots, u$, we can compute $\sum_{\hat{b}_j \in g_i} \hat{b}_j$ as $\mathbf{v}'_1[u] - \mathbf{v}'_1[l-1]$ and $\sum_{\hat{b}_j \in g_i} \hat{b}_j^2$ as $\mathbf{v}'_2[u] - \mathbf{v}'_2[l-1]$ in $O(1)$ time. Thus, calculating the cost of any group requires $O(1)$ time. Since there are $O(N^2)$ possible groups, we can compute all their costs in $O(N^2)$. Finally, in order to find the grouping strategy that minimizes $\hat{err}_2$, we employ the dynamic programming procedure of [17], which runs in $O(N^2)$ time. Therefore, our algorithm has total running time $O(N^2)$.

We conclude this section with an improvement on the accuracy yielded by the use of the squared error. Recall that our algorithm computes the group costs on the noisy histogram in order to ensure $\epsilon_2$-differential privacy. Thus, the grouping strategy that minimizes $\hat{err}_2$, may not minimize $err_2$ (defined on the original bins). In order to alleviate the effects of the extra noise in $\hat{err}_2$ we exploit the following observation. Using a similar approach as in the proof of Lemma 1 in [24], we can show that each group is expected to have its cost increased due to noise by $2\frac{|g_i|-1}{\epsilon_1^2}$, i.e., proportionally to the group size. The extra error leads to smaller groups for $\hat{err}_2$ minimization, compared to $err_2$. To mitigate this, we reduce the calculated cost $\hat{c}_i$ of each group $g_i$ by $2\frac{|g_i|-1}{\epsilon_1^2}$, before applying dynamic programming. Compared to its direct competitor [26], our algorithm improves the utility by up to 70% and the complexity by $N$.

## 5.2 Fixed Queries

The main idea is to use the *prefix sums* [15] as query workload for the fixed queries component. A prefix sum query over $\mathbf{h}$ is simply described by an index $j$, and returns the sum of bins $b_1, \ldots, b_j$, i.e., $\sum_{i=1}^{j} \mathbf{h}[i]$. There are $N$ prefix sums, hereafter represented by a vector $\mathbf{s}$ such that $\mathbf{s}[j] = \sum_{i=1}^{j} \mathbf{h}[i]$ for $j = 1, \ldots, N$. Moreover, observe that *any arbitrary* query can be always computed by the subtraction of *exactly two* prefix sums; for instance, interval $[i_l, i_u]$ is answered as $\mathbf{s}[i_u] - \mathbf{s}[i_l - 1]$.

The proposed method takes advantage of the fact that there are $N$ prefix sums, as opposed to $O(N^2)$ possible queries, to improve the complexity of the optimized matrix mechanism of DAWA by a factor of $N$. It considers the prefix sums as the fixed workload $\mathbf{W}$, and produces a noisy histogram $\hat{\mathbf{h}}$. The latter enables the computation of a vector of noisy prefix sums $\hat{\mathbf{s}}$, such that $\hat{\mathbf{s}}[i] = \sum_{i=1}^{j} \hat{\mathbf{h}}[i]$. Then, each interval query $[i_l, i_u]$ is computed in $O(1)$ as $\hat{\mathbf{s}}[i_u] - \hat{\mathbf{s}}[i_l - 1]$. Since the fixed queries component leads to highly accurate $\hat{\mathbf{s}}[i]$, the query result is expected to have very low error.

The time complexity of the method is $O(|\mathbf{W}|N \log N) = O(N^2 \log N)$, since now $|\mathbf{W}| = N$. The expected error is at most two times larger than that of DAWA's matrix mechanism because our method subtracts two noisy values from the prefix sums array to answer an interval query, while DAWA's matrix mechanism essentially returns a value for the same interval. However, in our experiments we demonstrate that the utility of our approach is practically the same when combined with the smoothing component.

# 6. NOVEL COMBINATIONS: HIERARCHICAL SMOOTHING

We seamlessly combine smoothing with the hierarchical component in order to benefit from the merits of both. Note that this combination can be used with either the average squared or absolute error, effectively yielding two different schemes. The running time is $O(N)$ when using the squared error, and $O(N \log N)$ with the absolute error metric.

Recall that the hierarchical component builds an aggregate tree in order to compose the query answer from a small number of noisy values, thus reducing the error resulting from noise aggregation as opposed to LPA. However, due to the increased sensitivity of the aggregate tree, it must add more noise per tree level than LPA. On the other hand, smoothing reduces the sensitivity of a set of bins via grouping and averaging, thus lowering the required noise. Our hierarchical smoothing method builds an aggregate tree similar to the hierarchical component (thus reducing the error from noise aggregation), but *smooths entire subtrees* via grouping and averaging similar to the smoothing component (thus reducing the per-level, per-bin noise).

Figure 1 illustrates the main idea. The scheme runs the smoothing *once* for the leaf level (i.e., for $\mathbf{h}$) with budget $\epsilon/4$, exploring groupings which consist of groups that only correspond to the leaves of *full* subtrees. Suppose that the black nodes in the figure comprise a group in the returned grouping strategy. We refer to the root of the subtree corresponding to a group as the *group root*. Next, the scheme creates the aggregate tree, *pruning* the nodes under the group roots (black nodes), and adds noise to the remaining nodes using budget $3\epsilon/4$. Finally, the scheme puts the pruned nodes back to the tree, deriving their values from their corresponding group root. Specifically, the value in the group root is distributed evenly across the nodes of the same level in the subtree. This is equivalent to smoothing the nodes at each level of the subtree via averaging.

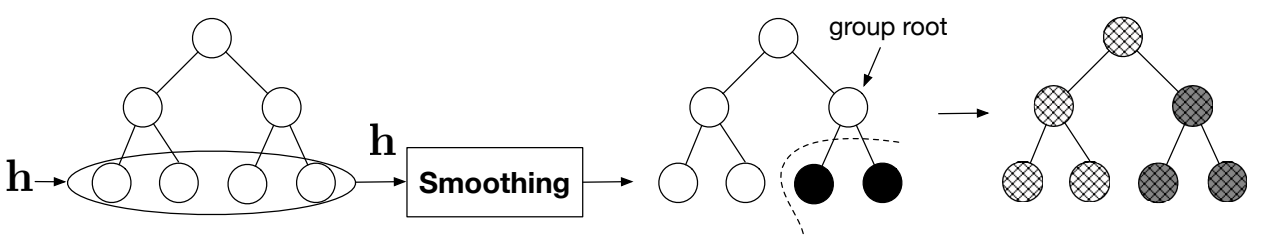

**Figure 1: Hierarchical Smoothing example**

The smoothing component spends budget $\epsilon/4$ and satisfies $\epsilon/4$-differential privacy. The hierarchical component spends budget $3\epsilon/4$ and satisfies $3\epsilon/4$-differential privacy-differential privacy. Moreover, both components work on non-disjoint inputs. Due to Theorem 2, the method satisfies $\epsilon/4 + 3\epsilon/4 = \epsilon$-differential privacy.

The running time of the scheme depends on the error metric. Observe that the number of groups examined by smoothing is equal to the number of nodes in the aggregate tree, i.e., $O(N)$. For the case of absolute error, the running time of smoothing is $O(N \log N)$, using the algorithm of [17]. For the squared error metric, the complexity is $O(N)$ using our optimal algorithm from Section 5.1. The hierarchical component runs in time linear in the number of input nodes, thus, in $O(N)$.

We can apply an additional utility optimization; instead of completely disregarding the nodes of a pruned subtree, we can actually utilize them to reduce the noise of its root. Specifically, for each level of the pruned subtree, we sum the node values and add noise, producing a noisy estimation of the root. Subsequently, we use the *average* of these estimations as the root noisy value. The mechanism then proceeds as described above, i.e., the root value is distributed evenly among the subtree nodes. This reduces the *squared* error of the root value by $t'$, where $t'$ is the height of the subtree.

Note also that ordering cannot be used before the smoothing because the final aggregate tree is built considering the order of the bins in $\mathbf{h}$. If ordering were used, grouping could select a group $g_i$, whose bins are not the leaves of a full subtree on $\mathbf{h}$ (since ordering may permute the bins of $\mathbf{h}$). Therefore, $g_i$ could not determine a group root to smooth a subtree, thus violating the scheme.

A final remark concerns allocation of budget $\epsilon$ to the various components. In this work, we followed the empirical allocation policies of the existing schemes. Determining the optimal allocation is out of our scope, but we consider it as an interesting problem for future work.

# 7. EXPERIMENTAL EVALUATION

Section 7.1 describes the datasets and the methodology. Sections 7.2 and 7.3 present our experimental results, and Section 7.4 summarizes them.

## 7.1 Methodology

We perform experiments with four real datasets, henceforth referred to as *Rome*[2], *Log* [14, 24, 25, 17], *Citations*

[2] http://apprendistato.informaservizi.it/

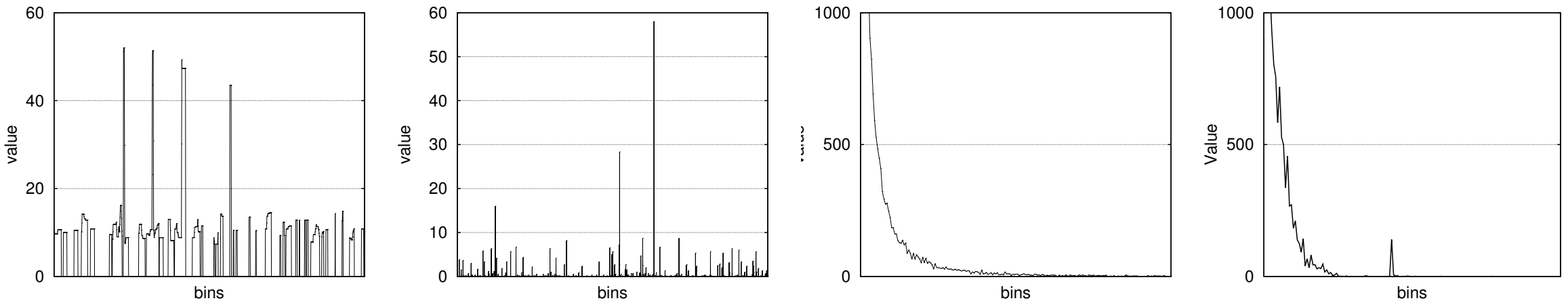

(a) *Rome* ($N = 1,442$) (b) *Log* ($N = 32,768$) (c) *Citations* ($N = 2,414$) (d) *PUMS* ($N = 7,578$)

**Figure 2: Data distributions**

**Table 1: Summary of methods**

| Scheme | Abbrv | Time |
|---|---|---|
| Smoothing with absolute error metric* | $\mathsf{S}_1$ | $O(N^2 \log N)$ |
| Approximate smoothing* | $\tilde{\mathsf{S}}$ | $O(N \log^2 N)$ |
| Smoothing with squared error metric** | $\mathsf{S}_2$ | $O(N^2)$ |
| Ordering and smoothing with absolute error metric | $\mathsf{O} + \mathsf{S}_1$ | $O(N^2 \log N)$ |
| Ordering and approximate smoothing | $\mathsf{O} + \tilde{\mathsf{S}}$ | $O(N \log^2 N)$ |
| Ordering and smoothing with squared error metric** | $\mathsf{O} + \mathsf{S}_2$ | $O(N^2)$ |
| Smoothing with absolute error metric and matrix mechanism* | $\mathsf{S}_1 + \mathsf{MM}$ | $O(N^3 \log N)$ |
| Approximate smoothing and matrix mechanism* | $\tilde{\mathsf{S}} + \mathsf{MM}$ | $O(N^3 \log N)$ |
| Smoothing with squared error metric and matrix mechanism | $\mathsf{S}_2 + \mathsf{MM}$ | $O(N^3 \log N)$ |
| Smoothing with absolute error metric and prefix matrix mechanism** | $\mathsf{S}_1 + \mathsf{PMM}$ | $O(N^2 \log N)$ |
| Approximate smoothing and prefix matrix mechanism** | $\tilde{\mathsf{S}} + \mathsf{PMM}$ | $O(N^2 \log N)$ |
| Smoothing with squared error metric and prefix matrix mechanism | $\mathsf{S}_2 + \mathsf{PMM}$ | $O(N^2 \log N)$ |
| Hierarchical* | $\mathsf{H}$ | $O(N)$ |
| Hierarchical Smoothing with absolute error metric | $\mathsf{S}_1 + \mathsf{H}$ | $O(N \log N)$ |
| Hierarchical Smoothing with squared error metric | $\mathsf{S}_2 + \mathsf{H}$ | $O(N)$ |

[11], and *PUMS*[3]. *Rome* consists of 1,442 bins, where each bin $b_i$ represents the number of cars on a specific road at time instance $i$. *Log* contains 32,768 bins with keyword statistics collected from Google Trends and American Online between 2004 and 2010. Each bin value corresponds to the frequency of a certain keyword. *Citations* contains information about scientific paper references. We create a histogram of 2,414 bins as in [21], where each bin $b_i$ is the number of papers cited $i$ times. Finally, *PUMS* is created from the Public Use Microdata Sample for California. It consists of 500,000 individuals, each having 10 attributes. We order the individuals according to the income attribute, resulting in a domain size of 7,578. *Rome*, *Log*, *Citations*, and *PUMS* have considerably different distributions, depicted in Figures 2(a), 2(b), 2(c), and 2(d), respectively. *Rome* exhibits high fluctuations at specific contiguous bins (reflecting the peak hours), and includes numerous small values (reflecting non-peak hours). *Log* consists of a large number of zero-valued bins along with a few fluctuations. *Citations* and *PUMS* are very sparse, and their consecutive bin values are similar, especially for bins that correspond to numerous citations (most such bins have zero values), or high incomes, respectively.

In our evaluation, we employ the state-of-the-art method for each component, and compare all possible combinations of the basic components, as summarized in Table 1. Techniques with an asterisk ("*") correspond to previous work, and are used as are. Methods with two asterisks ("**") have been proposed before, but utilize our improved smoothing with squared error metric presented in Section 5.1 or the fixed queries with the prefix sums as workload presented in Section 5.2. The rest correspond to novel combinations that have not been explored in the literature.

[3]https://www.census.gov/census2000/PUMS5.html

Specifically, $\mathsf{S}_1$ incorporates the smoothing algorithm of [17], based on the absolute error metric. $\tilde{\mathsf{S}}$ is the approximate smoothing of [17]. $\mathsf{S}_2$ applies smoothing using the squared error metric, and it utilizes the quadratic algorithm and utility optimization described in Section 5.1. $\mathsf{O}$ represents the ordering component, $\mathsf{MM}$ the matrix mechanism (fixed queries component) of [17] implemented with input all the possible interval queries, and $\mathsf{PMM}$ corresponds to the matrix mechanism with the prefix sums as workload (see Section 5.2). $\mathsf{O} + \mathsf{S}_2$ is the method of [26] with our $\mathsf{S}_2$ algorithm, while $\mathsf{S}_1 + \mathsf{MM}$ and $\tilde{\mathsf{S}} + \mathsf{MM}$ are essentially the methods presented in [17]. $\mathsf{H}$ implements the aggregate tree, using all optimizations of [21]. Finally, ordering cannot be used with methods that utilize the hierarchical or the fixed queries component, because these methods take advantage of the fact that interval queries are computed on consecutive bins. Ordering may permute the bins of the histogram, diminishing this advantage.

We distinguish five method groups (separated by horizontal lines in Table 1) based on component combinations; smoothing ($\mathsf{S}_1$, $\tilde{\mathsf{S}}$, $\mathsf{S}_2$), ordering and smoothing ($\mathsf{O} + \mathsf{S}_1$, $\mathsf{O} + \tilde{\mathsf{S}}$, $\mathsf{O} + \mathsf{S}_2$), smoothing and fixed queries ($\mathsf{S}_1 + \mathsf{MM}$, $\tilde{\mathsf{S}} + \mathsf{MM}$, $\mathsf{S}_2 + \mathsf{MM}$, $\mathsf{S}_1 + \mathsf{PMM}$, $\tilde{\mathsf{S}} + \mathsf{PMM}$, $\mathsf{S}_2 + \mathsf{PMM}$), hierarchical ($\mathsf{H}$), and hierarchical smoothing ($\mathsf{S}_1 + \mathsf{H}$, $\mathsf{S}_2 + \mathsf{H}$). In the following diagrams, methods that utilize smoothing with the absolute error metric are represented with a cross, approximate smoothing with a circle, and smoothing with the squared error metric with a triangle. Combinations of smoothing and fixed queries with prefix sums are in red color, smoothing and fixed queries with all possible queries as workload are in orange, ordering and smoothing are in blue, and hierarchical smoothing are in green color.

All methods were implemented in Java and executed on an Intel Core i7 2.5GHz with 16GB of RAM, running MacOS 10.12. For each method, we generated 100 histograms, ran the queries, and reported the average error. Methods utilizing the matrix mechanism (MM) terminated only in *Rome* and *Citations* due to their small domain sizes; therefore, MM is not included in the evaluation of the other datasets (it would take about a month to terminate on *Log*). In Section 7.2, we assess the running time versus the error, and identify the best method for each group. Then, in Section 7.3 we evaluate the error of the selected group representatives for different $\epsilon$ values and interval length $r$.

## 7.2 Running Time vs. Error

We initially assess two evaluation metrics: (i) the execution time for building the differentially private histogram, and (ii) the mean squared error (MSE). We apply two $\epsilon$ values, namely 0.01 (small $\epsilon$) and 1 (large $\epsilon$), and two query types (point and interval queries). Specifically, we generate all $N$ possible point queries, and measure the average MSE per bin. For the interval queries we generate all $\binom{N}{2} + N$ of them, and report the average MSE per query. Each plot depicts in bold the methods that are on the skyline for each experiment; the rest are dominated by some skyline method, i.e., they are inferior in terms of both efficiency and utility.

Figure 3 focuses on the *Rome* dataset for $\epsilon = 0.01$. As shown in Figure 3(a) for point queries, the skyline consists of $\tilde{S}$, $S_1 + H$, and $S_2 + H$, with $\tilde{S}$ achieving the highest utility and $S_1 + H$ being the most efficient for time critical applications. In general, smoothing methods offer low error because they mostly target point queries. On the other hand, since the fixed query component aims at range queries, it does not offer utility benefits with respect to the corresponding $\tilde{S}$, $S_1$ and $S_2$ (in some cases the utility degrades). Hierarchical methods ($H$, $S_1 + H$, $S_2 + H$) are very fast, but with the exception of $S_1 + H$, they yield rather high error. All ordering based methods ($O + S_1$, $O + \tilde{S}$, $O + S_2$) are dominated because the returned order of the bins is far from optimal due to the limited privacy budget.

For interval queries (Figure 3(b)) the skyline contains $S_1$, $S_2$, and $S_2 + H$. In general, due to the small number of bins $(1, 442)$, smoothing offers good utility without the need to combine it with other components because most interval queries are rather short, involving a few bins. Observe that methods based on prefix sums (e.g., $S_1 + PMM$) have similar error to those with workload all possible queries (e.g., $S_1 + MM$), while being orders of magnitude faster. Combinations involving the hierarchical component ($S_1 + H$, $S_2 + H$) are the most efficient, while offering high utility. Techniques based on ordering fail in this setting as well.

Figure 4 repeats the experiment of Figure 3 for $\epsilon = 1$. For point queries (Figure 4(a)) the skyline consists of $\tilde{S}$, $O + \tilde{S}$, and $H$. Compared to Figure 3(a), hierarchical methods do not take advantage of the increased privacy budget, and their relative utility with respect to the rest is worse. On the other hand, the relative performance of techniques that combine ordering and smoothing improves; finding the correct ordering before smoothing is important for the output error. For interval queries and large $\epsilon$ (Figure 4(b)), methods that integrate smoothing with hierarchical or fixed queries components perform better than those based solely on smoothing. In this setting, $S_1 + PMM$ and $S_2 + H$ are the dominant techniques. Observe that for large $\epsilon$, no method is on the skylines of both point and interval queries.

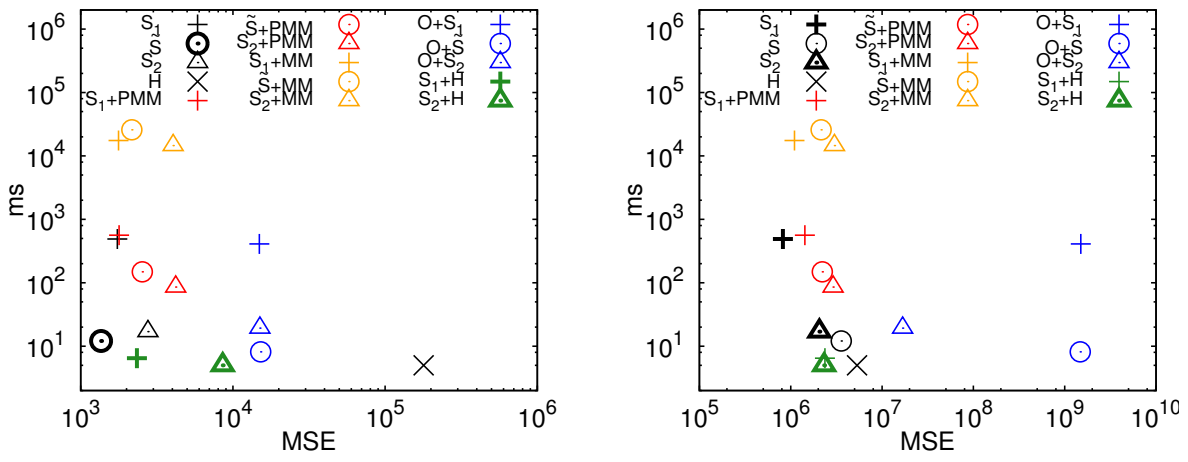

(a) Point Queries (b) Interval Queries

**Figure 3: Time vs. Error for small $\epsilon$ for *Rome***

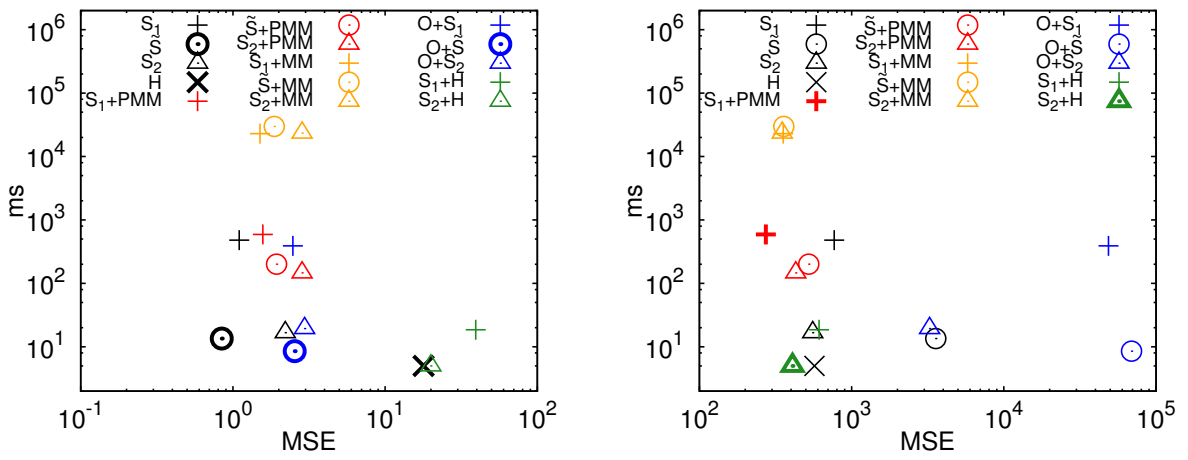

(a) Point Queries (b) Interval Queries

**Figure 4: Time vs. Error for large $\epsilon$ for *Rome***

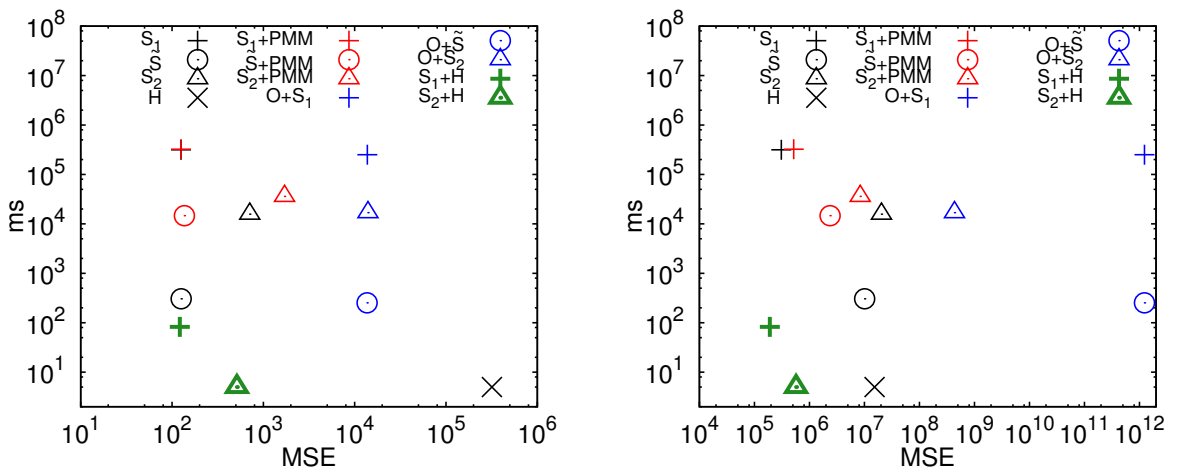

(a) Point Queries (b) Interval Queries

**Figure 5: Time vs. Error for small $\epsilon$ for *Log***

Figure 5 plots the skyline between the execution time and the MSE of each method for $\epsilon = 0.01$ and the *Log* dataset. The matrix mechanism with workload all possible queries failed to terminate in a reasonable time due to the numerous bins $(32, 768)$. According to Figures 5(a) and 5(b), $S_1 + H$ and $S_2 + H$ are the best methods for both point and interval queries, with $S_1 + H$ offering better utility at the expense of higher running time. Due to the large histogram size, the average range is long; thus, the fixed queries and hierarchical components yield utility benefits for interval queries. Smoothing methods are also competitive because when $\epsilon$ is small, they smooth more bins in order to maintain low error (due to noise addition).

Similarly, Figure 6 plots the results for $\epsilon = 1$ and the *Log* dataset. For point queries (Figure 6(a)), the skyline contains $\tilde{S}$, $O + \tilde{S}$ and $S_2 + H$, with $\tilde{S}$ offering the best utility and $S_2 + H$ the highest efficiency. The latter is also the absolute winner for interval queries in terms of both accuracy and running time (Figure 6(b)). Comparing with the case of $\epsilon = 0.01$ (Figure 5), methods that utilize smoothing with the absolute error metric offer better results than those with the squared error, when $\epsilon$ is small. This is due to the fact that, for the same $\epsilon$ value, $S_1$ needs less noise than $S_2$ to compute the grouping strategy. For large $\epsilon$ values, methods based on the squared error are in general better because they have enough privacy budget to discover the correct groups, and they target explicitly the metric used in the evaluation.

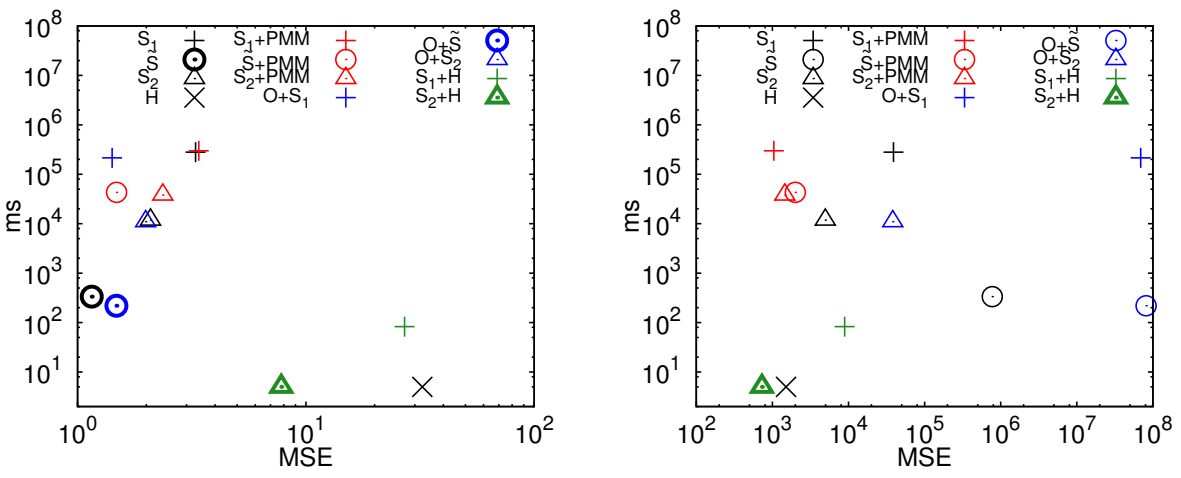

(a) Point Queries (b) Interval Queries

**Figure 6: Time vs. Error for large $\epsilon$ for *Log***

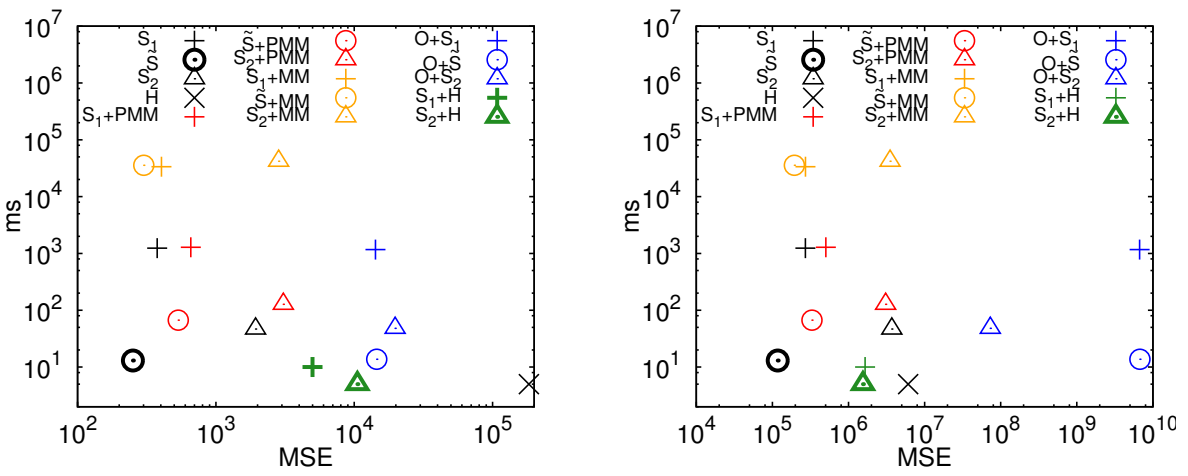

(a) Point Queries (b) Interval Queries

**Figure 7: Time vs. Error for small $\epsilon$ for *Citations***

Figure 7 depicts the execution time and the MSE for $\epsilon = 0.01$ and the *Citations* dataset. The *Citations* histogram is small ($2,414$ bins), implying that the average range is short. Moreover, as shown in Figure 2(c), consecutive bins have similar values (especially for bins corresponding to high citations, most of which are empty). Consequently, smoothing methods can easily generate large groups with low error, even if the privacy budget is small. $\tilde{\mathsf{S}}$ achieves the best utility in case of both point and interval queries. For point (resp. interval) queries, the skyline also contains $\mathsf{S}_1 + \mathsf{H}$ and $\mathsf{S}_2 + \mathsf{H}$ ($\mathsf{S}_2 + \mathsf{H}$), due to their efficiency, with however much larger error compared to $\tilde{\mathsf{S}}$.

Figure 8 repeats the experiment of Figure 7 for $\epsilon = 1$. Again, $\tilde{\mathsf{S}}$, $\mathsf{S}_1 + \mathsf{H}$, $\mathsf{S}_2 + \mathsf{H}$ are on the skyline for point queries (Figure 8(a)), with $\mathsf{S}_1 + \mathsf{H}$ and $\mathsf{S}_2 + \mathsf{H}$ offering utility closer to that of $\tilde{\mathsf{S}}$ due to the large privacy budget available in this setting. For interval queries (Figure 8(b)), the skyline consists of $\mathsf{S}_1$, $\tilde{\mathsf{S}} + \mathsf{PMM}$, $\mathsf{S}_1 + \mathsf{H}$ and $\mathsf{S}_2 + \mathsf{H}$. $\tilde{\mathsf{S}}$ is dominated in this setting because the privacy budget allows $\mathsf{S}_1$ to find a better grouping strategy. Moreover, the high $\epsilon$ value enhances the utility of methods that combine smoothing with either the fixed queries or hierarchical component when answering interval queries.

The last set of experiments focuses on the *PUMS* dataset, which is similar to *Citations*. As such, for point queries (Figure 9(a)), the skyline contains $\tilde{\mathsf{S}}$ and $\mathsf{S}_2 + \mathsf{H}$. However, *PUMS* incorporates bins with more abrupt value fluctuations than *Citations*. Thus, $\mathsf{O} + \tilde{\mathsf{S}}$ is also on the skyline because ordering successfully eliminates the value fluctuations of neighboring bins. In the case of interval queries (Figure 9(b)), the existence of zero-valued bins and the large histogram size ($7,578$) allows the effective combination of smoothing with the fixed queries or hierarchical component. Consequently, $\mathsf{S}_1 + \mathsf{PMM}$ and $\tilde{\mathsf{S}} + \mathsf{PMM}$ are on the skyline along with $\mathsf{S}_2 + \mathsf{H}$. The diagram for $\epsilon = 1$ and point queries (Figure 10(a)) is similar to the case of small $\epsilon$ (Figure 9(a)). However, as shown in Figure 10(b), for interval queries, $\mathsf{S}_2 + \mathsf{H}$ benefits more from the larger budget, and dominates all other methods in terms of both utility and efficiency.

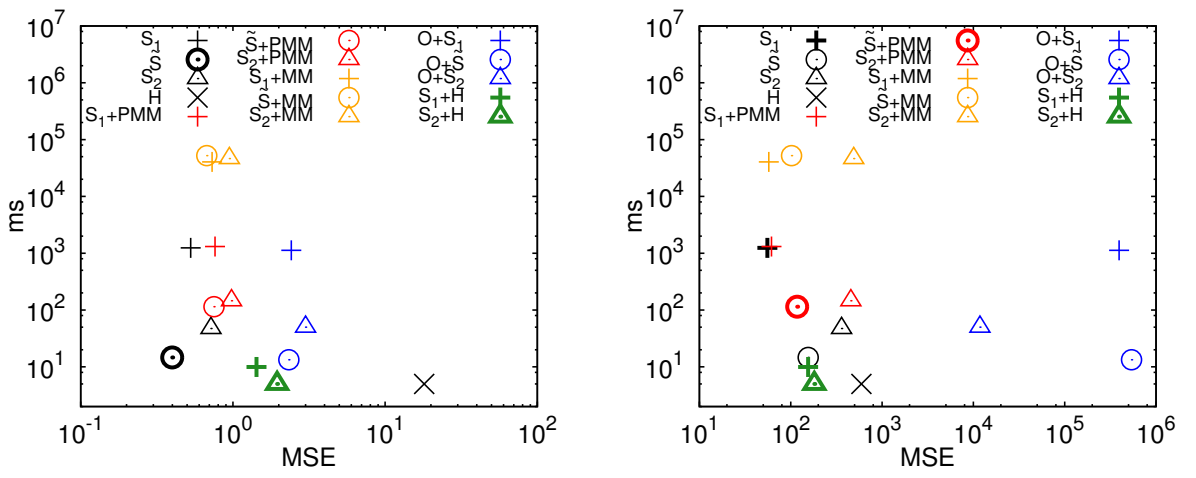

(a) Point Queries (b) Interval Queries

**Figure 8: Time vs. Error for large $\epsilon$ for *Citations***

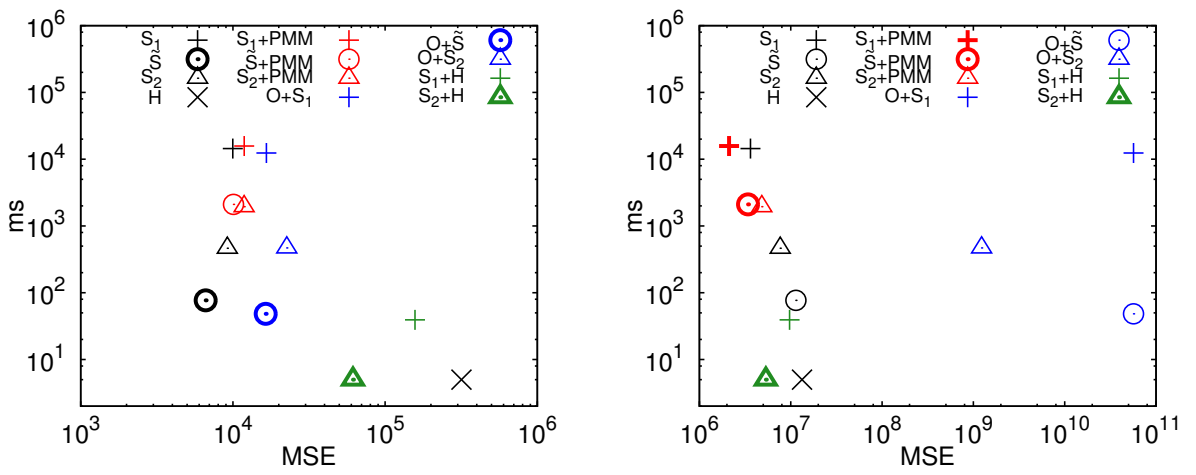

(a) Point Queries (b) Interval Queries

**Figure 9: Time vs. Error for small $\epsilon$ for *PUMS***

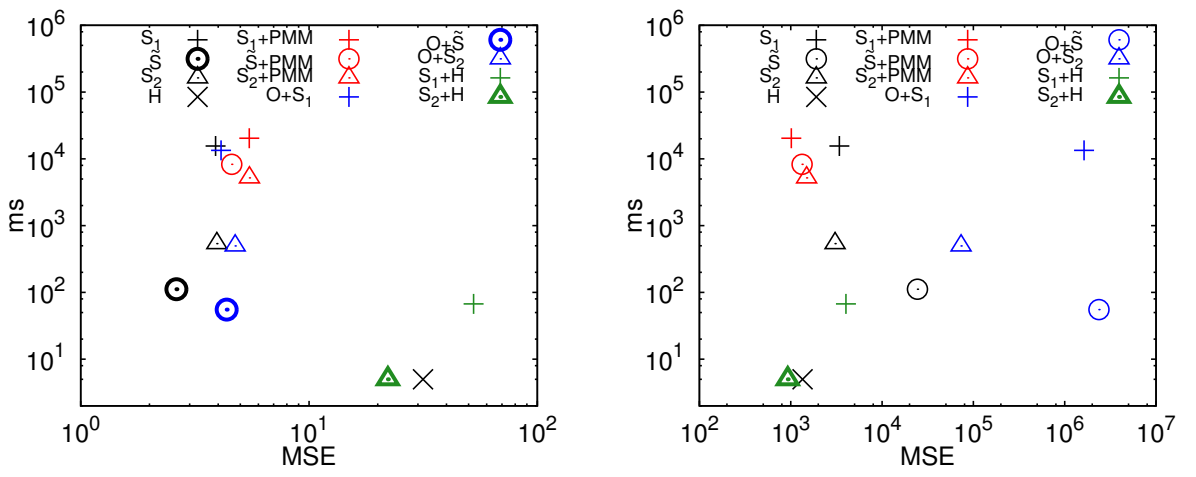

(a) Point Queries (b) Interval Queries

**Figure 10: Time vs. Error for large $\epsilon$ for *PUMS***

Table 2 summarizes the number of appearances of each method in some skyline for the Running Time vs. Error experiments. $\mathsf{S}_2 + \mathsf{H}$ offers the best trade-off since it is on the skyline in almost every setting, followed by $\tilde{\mathsf{S}}$ and $\mathsf{S}_1 + \mathsf{H}$. The worst methods are the combinations of smoothing and matrix mechanism with workload every possible query, failing to offer better utility than their prefix sums workload counterparts, while having worse time efficiency. For the combination of ordering and smoothing, only $\mathsf{O} + \tilde{\mathsf{S}}$ manages to be on skylines.

## 7.3 Utility Evaluation

In the sequel, we first identify the top method per component combination group. Recall that we have five groups: smoothing, ordering and smoothing, smoothing and fixed queries, hierarchical, and hierarchical smoothing. As top method per group, we select the one that lies on the most skylines per category, denoted with bold text in Table 2: $\tilde{\mathsf{S}}$ for smoothing, $\mathsf{O} + \tilde{\mathsf{S}}$ for ordering and smoothing, $\mathsf{S1} + \mathsf{PMM}$ for smoothing and fixed queries (although it is on the same

## Table 2: Methods on the skyline of Time vs. Error

| Method | Small $\epsilon$ Point queries *Rome* | *Log* | *Citations* | *PUMS* | Small $\epsilon$ Interval queries *Rome* | *Log* | *Citations* | *PUMS* | Large $\epsilon$ Point queries *Rome* | *Log* | *Citations* | *PUMS* | Large $\epsilon$ Interval queries *Rome* | *Log* | *Citations* | *PUMS* | Total |
|---|---|---|---|---|---|---|---|---|---|---|---|---|---|---|---|---|---|
| $\mathsf{S_1}$ | | | | | ✓ | | | | | | | | | | ✓ | | 2 |
| $\tilde{\mathsf{S}}$ | ✓ | | ✓ | ✓ | | | ✓ | | ✓ | ✓ | ✓ | ✓ | | | | | 8 |
| $\mathsf{S_2}$ | | | | | ✓ | | | | | | | | | | | | 1 |
| $\mathsf{O+S_1}$ | | | | | | | | | | | | | | | | | 0 |
| $\mathbf{O+\tilde{S}}$ | | | | ✓ | | | | | ✓ | ✓ | | ✓ | | | | | 4 |
| $\mathsf{O+S_2}$ | | | | | | | | | | | | | | | | | 0 |
| $\mathsf{S_1+MM}$ | | | | | | | | | | | | | | | | | 0 |
| $\tilde{\mathsf{S}}\mathsf{+MM}$ | | | | | | | | | | | | | | | | | 0 |
| $\mathsf{S_2+MM}$ | | | | | | | | | | | | | | | | | 0 |
| $\mathbf{S_1+PMM}$ | | | | | | | | ✓ | | | | | ✓ | | | | 2 |
| $\tilde{\mathsf{S}}\mathsf{+PMM}$ | | | | | | | | ✓ | | | | | | | ✓ | | 2 |
| $\mathsf{S_2+PMM}$ | | | | | | | | | | | | | | | | | 0 |
| **H** | | | | | | | | | ✓ | | | | | | | | 1 |
| $\mathsf{S_1+H}$ | ✓ | ✓ | ✓ | | | ✓ | | | | | ✓ | | | | ✓ | | 6 |
| $\mathbf{S_2+H}$ | ✓ | ✓ | ✓ | ✓ | ✓ | ✓ | ✓ | ✓ | | ✓ | ✓ | ✓ | ✓ | ✓ | ✓ | ✓ | 15 |

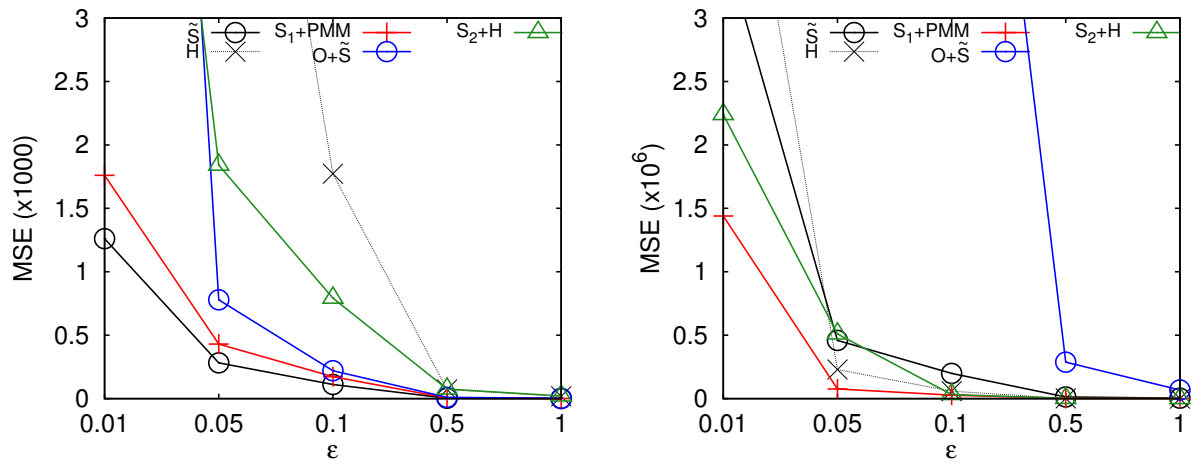

(a) Point Queries (b) Interval Queries

**Figure 11: Error vs. $\epsilon$ for *Rome***

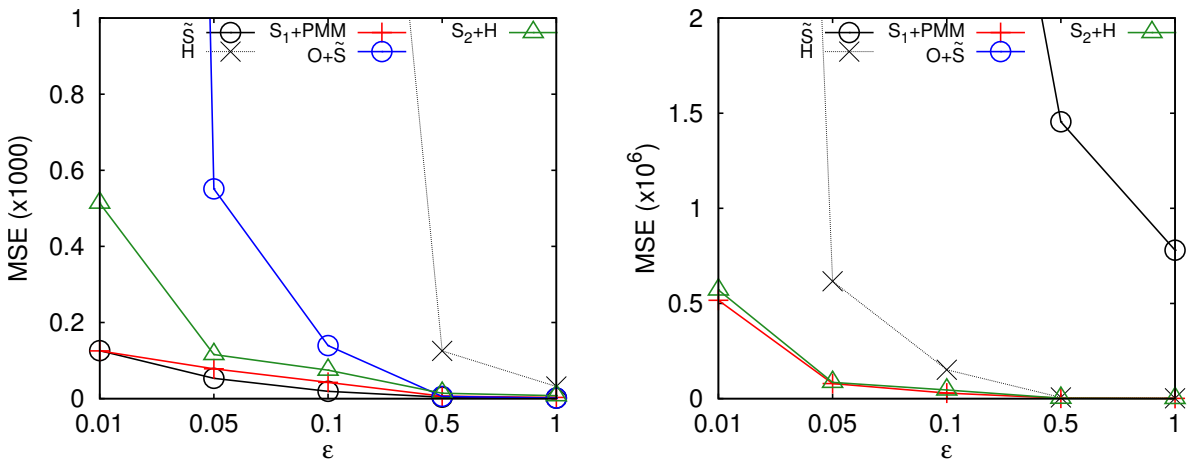

(a) Point Queries (b) Interval Queries

**Figure 12: Error vs. $\epsilon$ for *Log***

number of skylines as $\tilde{\mathsf{S}}+\mathsf{PMM}$, it offers better utility than the latter in most settings), $\mathsf{H}$ for hierarchical, and $\mathsf{S_2+H}$ for hierarchical smoothing. Then, we further evaluate the top methods in terms of output error for different $\epsilon$ values and interval query sizes. In the first set of experiments, we assess the MSE for privacy budget $\epsilon$ ranging from 0.01 to 1.

Figure 11 compares the best methods using *Rome*. For point queries (Figure 11(a)), $\tilde{\mathsf{S}}$ retains its low error even for small $\epsilon$ values because it merges more bins together in order to reduce the error due to the noise addition. $\mathsf{S1+PMM}$ also benefits from the smoothing component and its utility is comparable to that of $\tilde{\mathsf{S}}$. $\mathsf{O}+\tilde{\mathsf{S}}$ yields low error only for $\epsilon \geq 0.05$ because, for smaller values, the ordering is far from the optimal, resulting in almost arbitrary bin grouping. $\mathsf{H}$ and $\mathsf{S_2+H}$ deteriorate significantly as $\epsilon$ decreases, with $\mathsf{S_2+H}$ offering much better results than $\mathsf{H}$ due to the smoothing component. As shown in Figure 11(b) for interval queries, $\mathsf{S1+PMM}$, followed by $\mathsf{S_2+H}$, achieves the lowest error because the fixed queries/hierarchical components aim at interval queries. $\mathsf{H}$ outperforms $\tilde{\mathsf{S}}$ for $\epsilon \geq 0.05$. Finally, $\mathsf{O}+\tilde{\mathsf{S}}$ achieves acceptable utility only for $\epsilon \geq 0.5$, since it is mainly designed for point queries, and requires large $\epsilon$ values to find an ordering which is close to optimal.

Figure 12 repeats the experiment using *Log*. For points queries (Figure 12(a)), the top-3 methods are $\tilde{\mathsf{S}}$, $\mathsf{S_1+PMM}$ and $\mathsf{S_2+H}$, in this order. $\mathsf{O}+\tilde{\mathsf{S}}$ is consistently outperformed by $\mathsf{S_2+H}$ in this setting. $\mathsf{H}$ is by far the worst choice. For interval queries (Figure 12(b)), $\mathsf{S_1+PMM}$ and $\mathsf{S_2+H}$ are the best methods benefiting from the merits of both their components. $\mathsf{H}$ performs well only for $\epsilon \geq 0.05$. $\tilde{\mathsf{S}}$ and $\mathsf{O}+\tilde{\mathsf{S}}$ are the worst methods due to the long interval queries (recall that *Log* has the largest number of bins); $\mathsf{O}+\tilde{\mathsf{S}}$ does not appear in the diagram because its error is larger than the bounds of the plot.

Figure 13 assesses the utility versus $\epsilon$ for the *Citations* dataset. Again, standalone smoothing can better adapt to limited privacy budget, while the methods that combine it with the fixed queries or hierarchical component deteriorate faster as $\epsilon$ decreases. Even for interval queries (Figure 13(b)), $\tilde{\mathsf{S}}$ is still better than $\mathsf{S_1+PMM}$ and $\mathsf{S_2+H}$, although the *Citations* histogram is larger than that of *Rome*. This is due to the zero-valued bins in *Citations*, which help $\tilde{\mathsf{S}}$ to create large groups with low error. $\mathsf{O}+\tilde{\mathsf{S}}$ is competitive only for point queries, and it performs even worse than $\mathsf{H}$ in the case of interval queries.

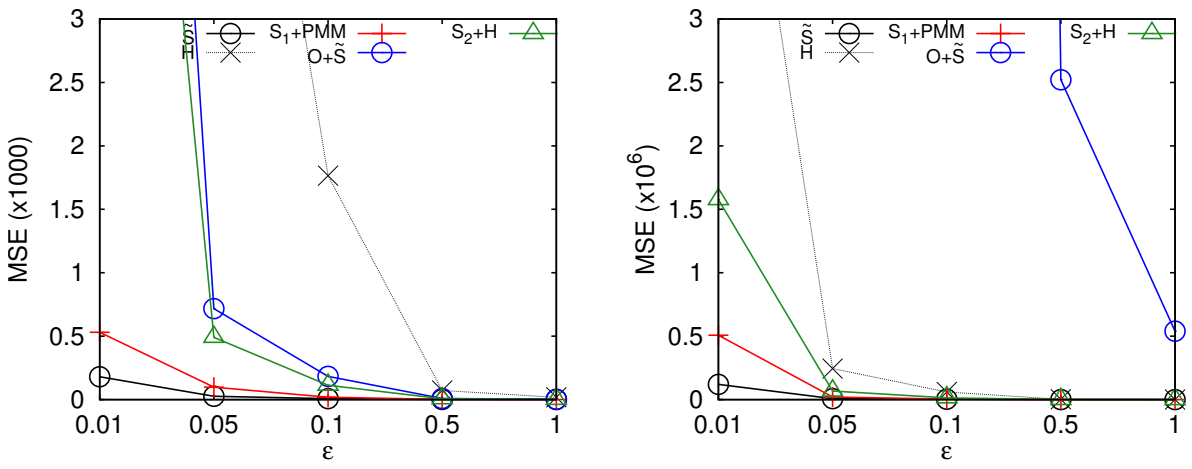

(a) Point Queries (b) Interval Queries

**Figure 13: Error vs. $\epsilon$ for *Citations***

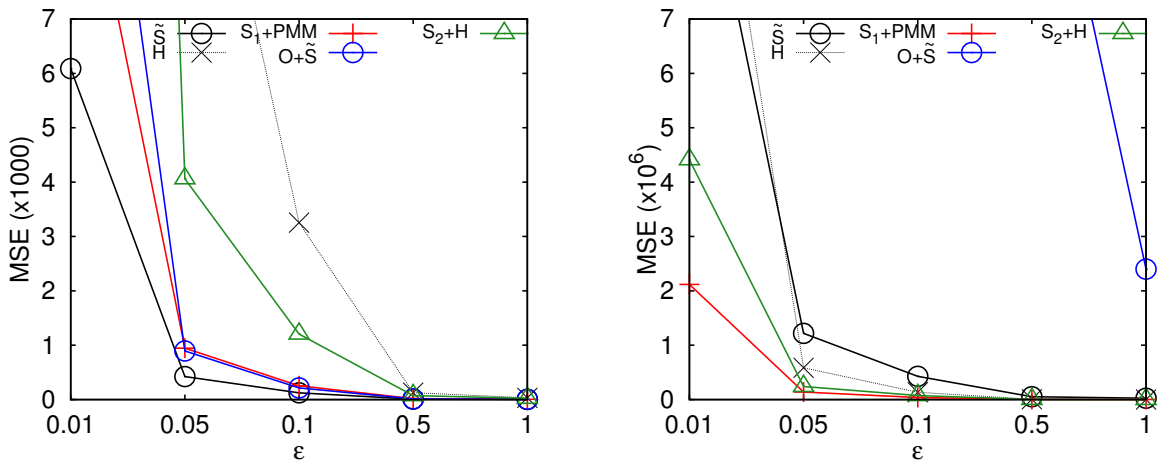

(a) Point Queries (b) Interval Queries

**Figure 14: Error vs. $\epsilon$ for *PUMS***

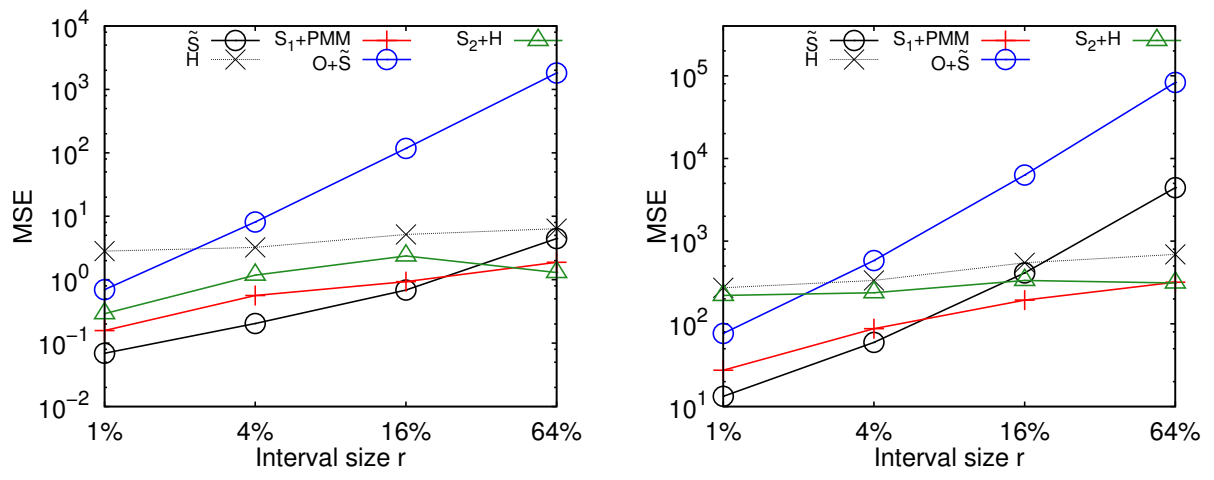

(a) Small $\epsilon$ (b) Large $\epsilon$

**Figure 15: Error vs. interval length $r$ for *Rome***

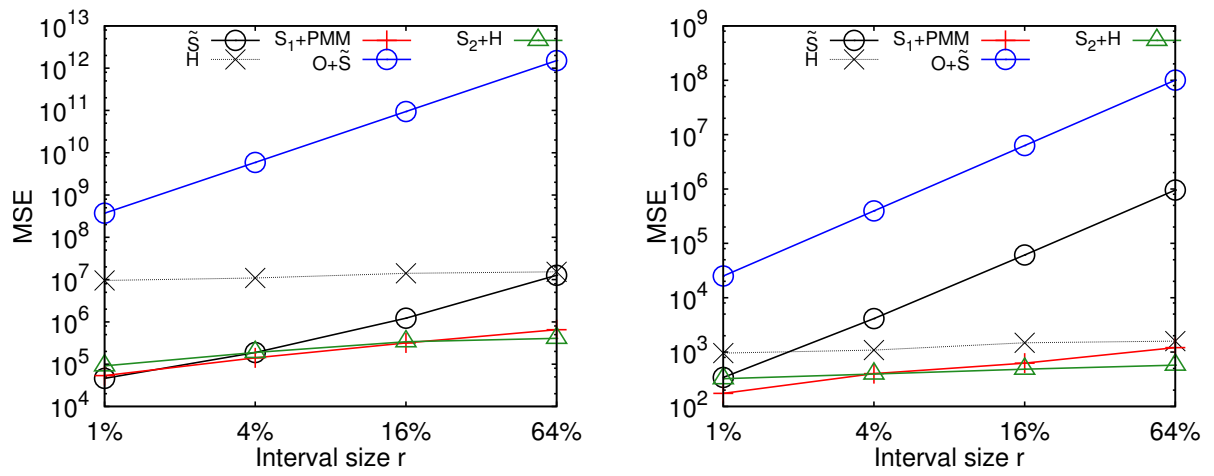

(a) Small $\epsilon$ (b) Large $\epsilon$

**Figure 16: Error vs. interval length $r$ for *Log***

Figure 14 repeats the experiment for *PUMS*. In case of point queries (Figure 14(a)), $\tilde{\mathsf{S}}$, $\mathsf{S}_1 + \mathsf{PMM}$, and $\mathsf{O} + \tilde{\mathsf{S}}$ behave similarly for different $\epsilon$ values and outperform both $\mathsf{S}_2 + \mathsf{H}$ and $\mathsf{H}$. This is due to the fact that the *PUMS* histogram is large with numerous zero-valued bins; consequently it is easy even for $\mathsf{O} + \tilde{\mathsf{S}}$ to correctly smooth these bins. The hierarchical component of $\mathsf{S}_2 + \mathsf{H}$ splits the privacy budget among tree levels that are not utilized for point queries. Thus it yields higher error, but it still outperforms $\mathsf{H}$. In case of interval queries (Figure 14(b)), $\mathsf{S}_1 + \mathsf{PMM}$ is the best method followed by $\mathsf{S}_2 + \mathsf{H}$ because the large size of the *PUMS* takes advantage of the fixed queries and hierarchical component to lower the error for long intervals.

Next, we evaluate the MSE for interval query sizes varying the query interval length $r$ from 1% to 64% of the histogram size. Figure 15 plots the MSE versus $r$ for *Rome*. In case of $\epsilon = 0.01$ (Figure 15(a)), $\tilde{\mathsf{S}}$ is very accurate for interval lengths up to 16%. For longer intervals, utilizing the fixed queries or hierarchical component in addition to smoothing, offers better results. $\mathsf{S}_1 + \mathsf{PMM}$ and $\mathsf{S}_2 + \mathsf{H}$ achieve higher utility than $\mathsf{H}$, even for very long intervals, demonstrating the benefits of combining components. For large $\epsilon$ values (Figure 15(b)), the results are similar, but in this case, the fixed queries and hierarchical components benefit more than simple smoothing from the large budget. For long intervals, even $\mathsf{H}$ eventually has lower error than $\tilde{\mathsf{S}}$. In general, $\mathsf{O} + \tilde{\mathsf{S}}$ is the worst method, except for short intervals.

Figure 16 illustrates the MSE versus the interval length $r$ for the *Log* dataset. For $\epsilon = 0.01$ (Figure 16(a)), we observe analogous results to the corresponding *Rome* plot (Figure 15(a)). However, the absolute intervals are much longer in this case due to the large histogram size (32,768 bins in *Log* versus 1,442 bins in *Rome*). Consequently, $\mathsf{S}_1 + \mathsf{PMM}$ and $\mathsf{S}_2 + \mathsf{H}$ have similar utility to $\tilde{\mathsf{S}}$ for short intervals, and outperform it when $r > 4\%$. For $\epsilon = 1$, $\mathsf{S}_1 + \mathsf{PMM}$ and $\mathsf{S}_2 + \mathsf{H}$ are better than $\tilde{\mathsf{S}}$ for all lengths. $\mathsf{H}$ performs well only for large privacy budget, while $\mathsf{O} + \tilde{\mathsf{S}}$ is not competitive.

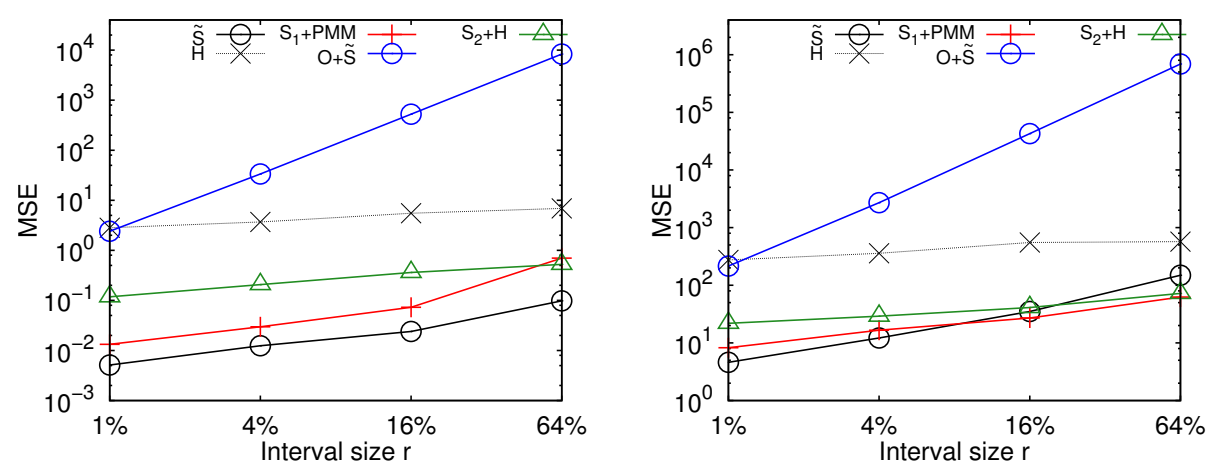

(a) Small $\epsilon$ (b) Large $\epsilon$

**Figure 17: Error vs. interval length $r$ for *Citations***

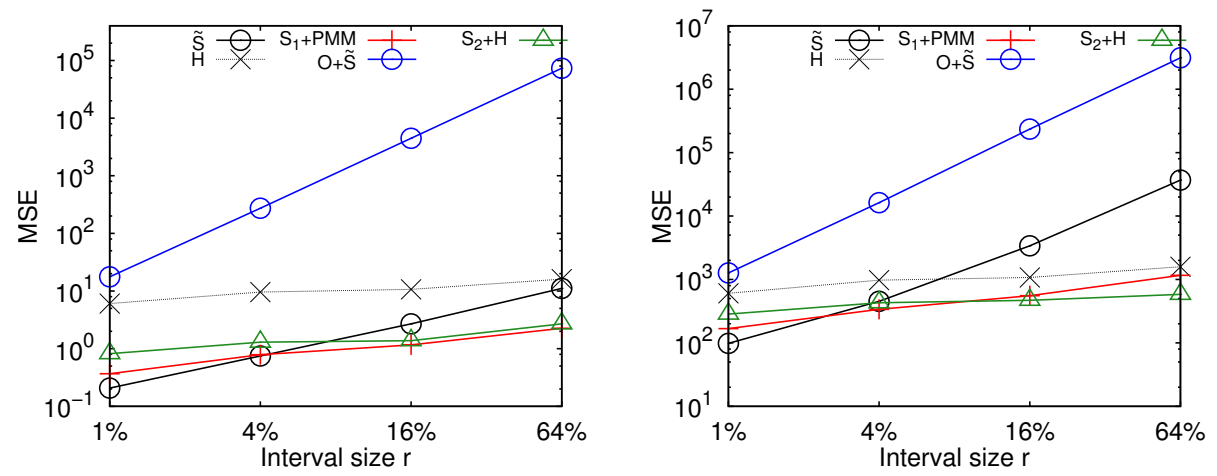

(a) Small $\epsilon$ (b) Large $\epsilon$

**Figure 18: Error vs. interval length $r$ for *PUMS***

Next we evaluate the utility of the methods on the *Citations* dataset. As shown in Figure 17(a), $\tilde{\mathsf{S}}$ is the winner for all interval lengths when $\epsilon = 0.01$, due to the small size and the high number of zero-valued bins of the histogram. Methods that add components distribute the privacy budget without any benefits, compared to standalone smoothing. For large $\epsilon$ (Figure 17(b)), the performance gap closes and eventually $\mathsf{S}_1 + \mathsf{PMM}$, $\mathsf{S}_2 + \mathsf{H}$ achieve higher utility than $\tilde{\mathsf{S}}$ when $r = 64\%$. $\mathsf{H}$ and $\mathsf{O} + \tilde{\mathsf{S}}$ fail for both privacy budgets.

Finally, Figure 18 focuses on *PUMS*. Although, similarly to *Citations*, *PUMS* has numerous zero-valued bins, it also exhibits larger value fluctuations, which are harder to group together, and it is much larger. Accordingly, the benefits of the hierarchical and fixed query components are substantial and the plots resemble those of *Log* (Figure 16). $\mathsf{S}_1 + \mathsf{PMM}$, $\mathsf{S}_2 + \mathsf{H}$ reach the utility of $\tilde{\mathsf{S}}$ when $r = 4\%$ and outperform it for longer intervals, even in the presence of low budget.

## 7.4 Takeaways

According to our evaluation, there is no overall winner, but the method of choice depends on the (i) query length, (ii) data characteristics, (iii) budget availability, (iv) efficiency requirements. Specifically for (i), point or short interval queries favor standalone smoothing. Methods integrating the fixed query component are more appropriate for queries of medium length, while the hierarchical component is the best choice for long queries, as it is the least sensitive to the interval length. Regarding (ii), small histograms where neighboring bins have similar values are ideal for standalone smoothing; otherwise, the fixed query or hierarchical components yield important benefits, especially for long queries. About (iii), simple smoothing is preferable in the presence of a limited privacy budget because this can be devoted entirely to grouping. Large budgets can improve utility through budget distribution to different components. Finally regarding (iv), hierarchical methods are the most efficient for time critical applications.

In terms of concrete methods as evaluated in Section 7.3:

1. $\tilde{\mathsf{S}}$ is the method of choice for point and short interval queries, especially for histograms where consecutive bins have similar values. Compared to the other standalone smoothing techniques ($\mathsf{S}_1$ and $\mathsf{S}_2$), it is faster ($O(N \log^2 N)$), and usually achieves lower error.

2. $\mathsf{S1} + \mathsf{PMM}$ has the most balanced behavior for all interval lengths (i.e., for both short and long ranges). However, it requires more privacy budget than $\tilde{\mathsf{S}}$ in order to be effective, and it is slower. Compared to the other techniques that combine smoothing with the fixed queries component, it has the lowest complexity ($O(N^2 \log N)$), and usually exhibits the highest utility.

3. $\mathsf{S}_2 + \mathsf{H}$ is the winner for long interval queries and the method of choice for time critical applications. Similar to $\mathsf{S1} + \mathsf{PMM}$, it requires sufficient privacy budget to achieve the best results. With linear complexity, it is faster than $\mathsf{S}_1 + \mathsf{H}$ and is the technique with most skyline appearances in the experiments of Section 7.2.

4. $\mathsf{H}$ and $\mathsf{O} + \tilde{\mathsf{S}}$ were outperformed by the above methods in all our settings. Specifically, although $\mathsf{H}$ implements all optimizations of [21], it is always worse than $\mathsf{S}_1 + \mathsf{H}$, suggesting that the hierarchical component is most effective when combined with smoothing. With few exceptions, $\mathsf{O} + \tilde{\mathsf{S}}$ yields the highest error, independently of the privacy budget, data and query characteristics.

# 8. CONCLUSION

This is the first evaluation of techniques for differentially private histograms that decomposes existing methods into basic components, namely smoothing, ordering, fixed queries, and hierarchical. This modular approach leads to novel optimization opportunities, including a smoothing algorithm based on the squared error, which improves utility and reduces the running time of the current state-of-the-art, and a prefix-sums approach that optimizes the fixed queries component. Moreover, it facilitates the development of original techniques, such as hierarchical smoothing, through innovative combinations of components. Our experimental evaluation assesses the utility and efficiency of all possible component combinations on four datasets with diverse characteristics. Although there no clear winner under all settings, some of the proposed optimizations/combinations are usually the methods of choice, demonstrating the benefits of our modular approach.